\documentclass{aa}
\usepackage{latexsym}
\usepackage{times}
\usepackage{graphicx}
\usepackage{natbib}

\bibpunct{(}{)}{;}{}{}{,}

\begin{document}

\thesaurus{07 
  ( 08.01.1; 08.01.3; 08.05.3; 08.16.4; 08.23.1; 13.21.5)}
  
\title{The iron abundance in hot central stars of planetary nebulae derived
  from IUE spectra\thanks {Based on observations with the International
    Ultraviolet Explorer (IUE) satellite}}
 
\subtitle{}

\author{J.L.\ Deetjen \and S. Dreizler \and T. Rauch \and K. Werner}

\offprints{J.L.\ Deetjen}

\mail{deetjen@astro.uni-tuebingen.de}

\institute{Institut f\"ur Astronomie und Astrophysik, 
  Universit\"at T\"ubingen, 
  Waldh\"auser Str. 64, 
  D--72076 T\"ubingen, Germany 
  }

\date{Received date; accepted date}
 
\titlerunning{The iron abundance in hot central stars of planetary nebulae}

\authorrunning{J.L.~Deetjen {et~al.}}

\maketitle

%
\begin{abstract}
  We present the first attempt to determine the iron abundance in hot
  central stars of planetary nebulae. We perform an analysis with fully
  metal-line blanketed NLTE model atmospheres for a sample of ten stars
  ($T_{\rm eff}$ \mbox{$\stackrel{>}{\mbox{\tiny $\sim$}}$} 70\,000\,K) for
  which high-resolution UV spectra are available from the IUE archive.  In
  all cases lines of \ion{Fe}{vi} or \ion{Fe}{vii} can be identified.  As a
  general trend, the iron abundance appears to be subsolar by 0.5--1~dex,
  however, the S/N of the IUE spectra is not sufficient to exclude a solar
  abundance in any specific case.  Improved spectroscopy by either FUSE or
  HST is necessary to verify the possibility of a general iron deficiency
  in central stars. The suspected deficiency may be the result of
  gravitational settling in the case of three high-gravity objects. For the
  other stars with low gravity and high luminosity dust fractionation
  during the previous AGB phase is a conceivable origin.  
\end{abstract}

%
%
%
\begin{figure}
  \resizebox{\hsize}{!}{\includegraphics{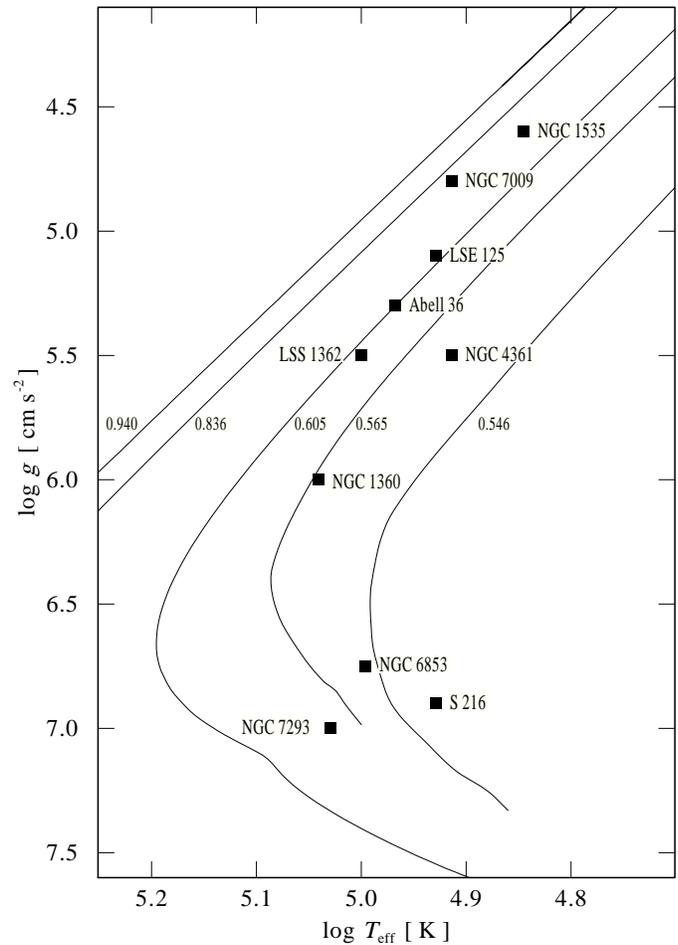}} 
  \caption{Location of the program stars in the $\log T_{\rm eff}$ -- $\log
    g$ diagram compared with post-AGB evolutionary tracks
    \citep{schoenberner:83a, bloecker:90a} for H-burning stars with
    different remnant masses (in M$_{\sun}$) as indicated.}
  \label{fig:tracks}
\end{figure}  

\begin{table*}
  \caption{
    The program stars and their atmospheric parameters. Values for $T_{\rm
    eff}$ [K], $\log g$ [cm s$^{-2}$],  and helium abundance were taken from
    literature. Columns 5 and 6 list our results for the Fe abundance and 
    the \ion{H}{i} column density along the line of sight as derived from
    the Ly$\alpha$ profile; the references are listed in
    Tab.~\ref{tab:references}  
    }  
  \label{tab:objects} 
  \begin{tabular}{lrrcccll}
    \hline
    \noalign{\smallskip}
    PN & $T_{\rm eff}$ & $\log g$ & n$_{\rm
    He}$/n$_{{\rm He}_{\sun}}$ & log n$_{\rm Fe}$/n$_{{\rm Fe}_{\sun}}$ &
    n$_{\ion{H}{i}}$ [cm$^{-2}$] & references & SWP numbers of co-added
    spectra \\   
    \noalign{\smallskip}
    \hline
    \noalign{\smallskip}
    \object{Abell\,36} &  93\,000 & 5.3  & 1.5 & $-$1.3 &
    $1.9\cdot10^{19}$ & 4, 5, 8, 9       & 16478,\quad 38132,\quad 47814,\quad 47815 \\
    
    \object{LSE\,125}  &  85\,000 & 5.1  & 0.5 & $-$0.5 &
    $3.0\cdot10^{19}$ & 2, 5             & 30269 \\

    \object{LSS\,1362} & 100\,000 & 5.5  & 1.0 & $-$0.5 &
    $1.6\cdot10^{19}$ & 3, 4, 5, 8       & 19830 \\

    \object{NGC\,1360} & 110\,000 & 6.0  & 1.0 & $-$0.5 &
    $3.0\cdot10^{18}$ & 1, 2, 4, 5, 8, 10 & 55902,\quad 56037,\quad 56038 \\

    \object{NGC\,1535} &  70\,000 & 4.6  & 1.2 & $-$0.5 &
    $1.4\cdot10^{19}$  & 1, 2, 4, 5, 8   & 56065,\quad 56066 \\

    \object{NGC\,4361} &  82\,000 & 5.5  & 0.5 & $-$1.3 &
    $5.0\cdot10^{18}$ & 1, 2, 4, 5, 8, 9 & 20440 \\

    \object{NGC\,6853} &  99\,000 & 6.8  & 1.0 & $-$0.5 &
    $3.0\cdot10^{18}$ & 5, 7, 8          & 18340 \\

    \object{NGC\,7009} &  82\,000 & 4.8  & 0.5 & $-$0.5 &
    $1.5\cdot10^{19}$ & 2, 4, 5, 8       & 05174,\quad 08580,\quad 08680,\quad 23383 \\

    \object{NGC\,7293} & 107\,000 & 7.0  & 0.5 & $-$0.5 &
    $5.5\cdot10^{18}$ & 4, 9             & 19859,\quad 48055,\quad 48063 \\

    \object{S\,216}    &  85\,000 & 6.9  & 0.1 & $-$0.5 &
    $5.0\cdot10^{18}$ & 4, 6, 7, 8       & 56071,\quad 56072 \\

    \noalign{\smallskip}    
    \hline
  \end{tabular}
\end{table*}

\section{Introduction} 
The continuing spectroscopic study of central stars of planetary nebulae
(CSPN) is motivated by two observational facts which reveal our incomplete
understanding of post-AGB stellar evolution. The first concerns the mere
existence of hydrogen-deficient objects which comprise the spectroscopic
types [WC] and PG\,1159. The second concerns the apparent lack of
hydrogen-rich CSPN at the hot end of the white dwarf cooling sequence,
suggesting that during this stage all H-rich objects become H-deficient.
The first problem initialized an almost complete analysis of all known
H-deficient post-AGB stars, resulting in accurate determinations of the
effective temperature ($T_{\rm eff}$) and surface gravity ($g$), as well as
the abundance of the main atmospheric constituents (He, C, O) and in some
cases trace of elements (H, N, Ne). The second problem could be solved by
the systematic search for \mbox{(pre-)} white dwarf central stars.
Subsequent analyses of the H-rich objects revealed $T_{\rm eff}$, $\log g$,
and H:He ratio and the results complement the analyses of other
H-rich CSPN performed in numerous previous analyses (see
Tab.~\ref{tab:references}).

Almost no effort has been yet made to analyze metal abundances in the
H-rich CSPN.  Interesting insight into AGB evolutionary phases may be
gained by such analyses, because obvious abundance anomalies exist among
stars with otherwise similar atmospheric parameters \citep{mendez:91a}.
Before embarking on such a project and discussing the abundance patterns it
is reasonable to obtain information about the primordial metallicity of
these objects by analyzing their iron abundance. The reason herefore is
that iron is not affected by nuclear processes during previous evolutionary
phases. Such an analysis requires UV spectroscopy. 
As a first step it is mandatory to check, if the IUE Final Archive contains
useful data. It is the aim of this paper to exploit that archive exhaustively
in order to identify and analyze quantitatively iron lines in hot central stars.

%
\begin{table}
  \caption{References in Tab.~\ref{tab:objects}}
  \label{tab:references}
    \begin{tabular}{c l c l}
    1 & \cite{mendez:81}      & 2  & \cite{mendez:88a}    \\
    3 & \cite{heber:88a}      & 4  & \cite{feibelman:90a} \\
    5 & \cite{mendez:91a}     & 6  & \cite{tweedy:92a}    \\
    7 & \cite{napiwotzki:phd} & 8  & \cite{tweedy:93a}    \\
    9 & \cite{quigley:93a}    & 10 & \cite{hoare:96a}     \\
  \end{tabular}   
\end{table}

\section{Sample selection and IUE data reduction}
Our sample selection started from a CSPN compilation by \cite{mendez:91a}
and we complemented his list of H-rich objects by several objects dealt
with in the more recent literature (e.g. see Tab.~\ref{tab:references}).
For these stars we checked the IUE archive for high resolution SWP spectra
and found data for 27 H-rich CSPN. From this sample those objects were
excluded, which have high mass-loss rates, i.e., those showing Of-type
optical spectra and those with extraordinarily strong P~Cygni resonance
line profiles in the UV. Thereby the final sample consisted of ten stars,
which can be analyzed reliably with our static model atmospheres. All
program stars have $T_{\rm eff}$ \mbox{$\stackrel{>}{\mbox{\tiny $\sim$}}$}
70\,000\,K.  Fig.~\ref{fig:tracks} shows their position in the $\log T_{\rm
  eff} - \log g$-diagram and Tab.~\ref{tab:objects} summarizes their
atmospheric parameters.

\begin{figure}
  \resizebox{\hsize}{!}{\includegraphics{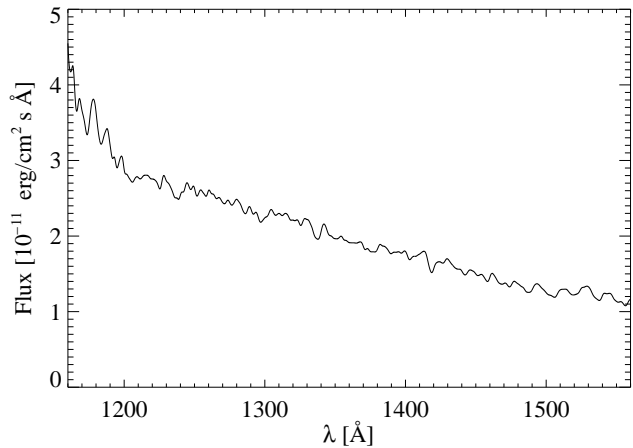}} 
  \caption{Calculated continuum of a typical co-added IUE spectrum}
  \label{fig:continuum}
\end{figure}

The SWP spectra were extracted from the IUE Final Archive and subject to
several reduction steps before comparison with model spectra. These steps
comprise the assembly of single Echelle orders in order to obtain a
complete spectrum in the 1200 -- 1600\,\AA\ range; elimination of the
strongest interstellar absorption lines by Voigt profile fitting
(Ly$\alpha$ and \ion{Si}{ii} lines); correction for radial velocity;
normalization of the spectra. The last step is rather delicate. Firstly the
ripple correction for the Echelle orders in the archival data is far from
being perfect (see Fig.~\ref{fig:continuum}) and has to be corrected.
Secondly, fixing the continuum is a rather subjective procedure and thus
the major error source for the quantitative analysis, because of the
generally high noise level in the spectra. To determine the continuum we
smoothed the spectrum with a median filter of 2\,\AA\ width, followed by a
convolution with a Gaussian profile of 2\,\AA\ FWHM. This method is
inapplicable in regions with P~Cygni profiles and the continuum has to be
determined manually. Furthermore the determined continuum around deep ISM
lines is too low and has to be raised manually.

%
\begin{figure}
  \resizebox{\hsize}{!}{\includegraphics{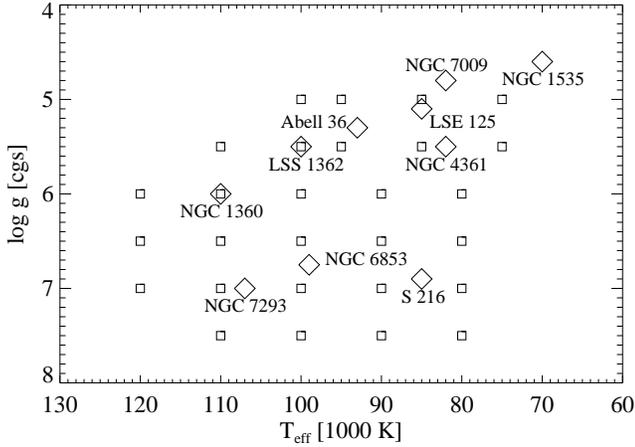}} 
  \caption{Distribution of the analyzed CSPN ($\Diamond$)
    and models ($\Box$) in the $T_{\rm eff}$~-~$\log g$~-~plane. For each
    point of the grid spectra for eight different Fe and Ni abundances have
    been calculated}
  \label{fig:grid}
\end{figure}

\begin{figure}
  \resizebox{\hsize}{!}{\includegraphics{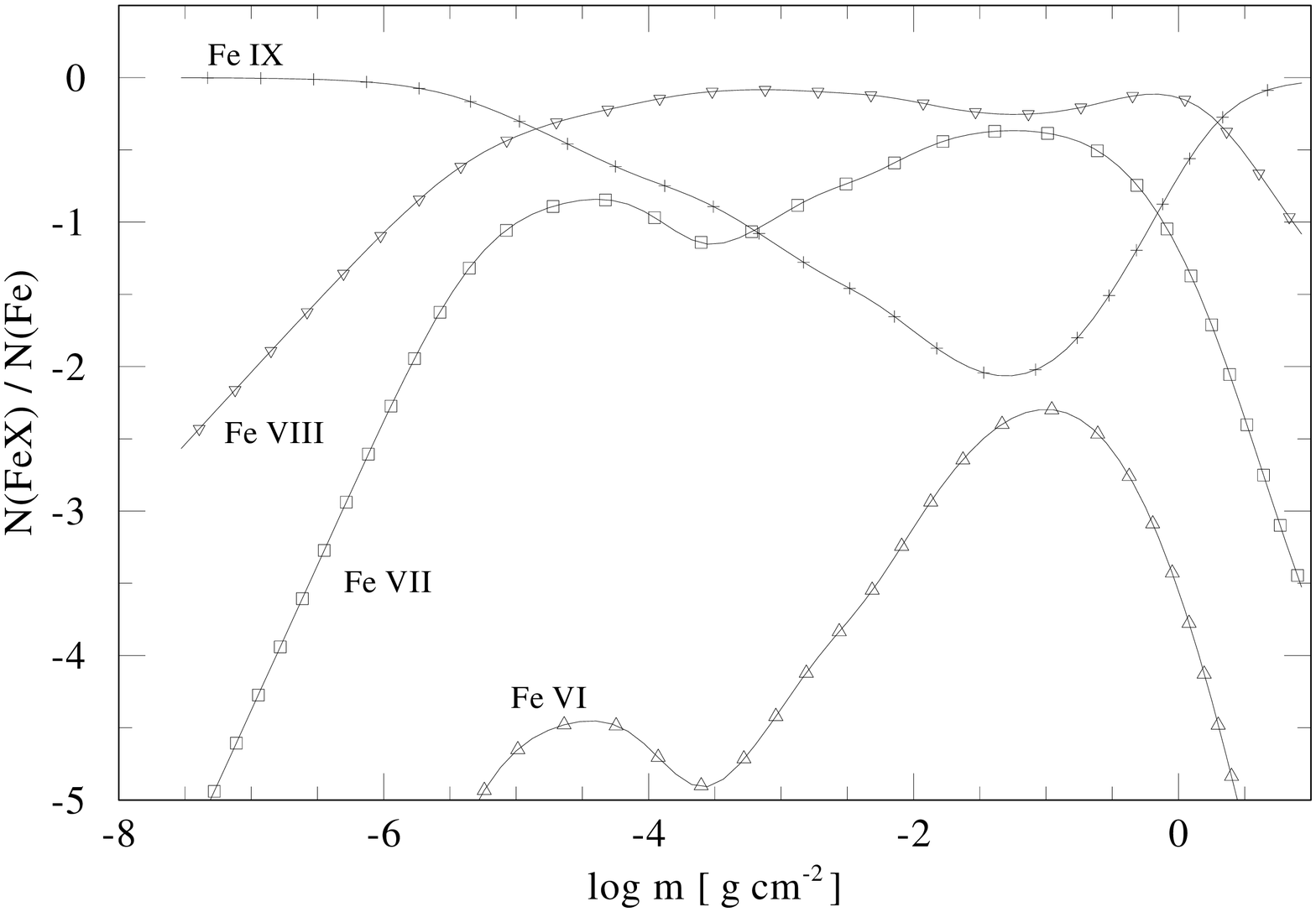}} 
  \caption{Ionization structure of iron in a model atmosphere with  $T_{\rm
      eff}$ = 110\,000 K, $\log g$ = 6.0 and solar Fe abundance. Note
    particularly the relative occupation of \ion{Fe}{vii} and \ion{Fe}{viii}    
    in the deeper layers of the atmosphere where the iron lines are formed
    (log\,$m=-2\ldots$0)}
  \label{fig:ionplot}
\end{figure}

\section{Model Atmospheres}
We have calculated a grid of plane-parallel non-LTE model atmospheres
(Fig.~\ref{fig:grid}) in radiative and hydrostatic equilibrium. The
computer code is based on the Accelerated Lambda Iteration method
\citep{werner:85a,werner:86a} and it can handle the line blanketing of
iron group elements by a statistical approach using superlevels and
superlines with an opacity sampling technique \citep{anderson:85a,
  dreizler:93a}. The latest version of the code and a detailed description
of the numerical method can be found in \cite{werner:99a}. Our model
atmospheres include the CNO elements as well as iron and nickel
self-consistently, i.e.\ their back-reaction on the atmospheric structure
is accounted for. In essence, they are very similar to the models described
in detail by \cite{haas:96a, haas:phd}, so we restrict ourselves here to a
summary of the model atoms in Tab.~\ref{tab:atoms}. The main extension of
that models refers to a detailed model atom for \ion{Fe}{viii}. This turned
out to be essential in order to reliably compute \ion{Fe}{vii} lines,
because the relative occupation of both ionization stages sensitively
depends on the detailed treatment of their respective level populations via
ionization and recombination processes.

As an example, the iron ionization structure in a typical model atmosphere
is shown in Fig.~\ref{fig:ionplot}.

%
\begin{table}
  \caption{
    Summary of model atoms used in our model atmosphere
    calculations. Numbers in brackets denote individual levels and lines
    used in the statistical NLTE line blanketing approach for iron and
    nickel. The model atom for each chemical element is closed by a single
    level representing the highest ionization stage (not listed
    explicitly)}  
  \label{tab:atoms}

  \begin{tabular}{l l r r r r}
    \hline
    \noalign{\smallskip}
    element & ion & \multicolumn{2}{l}{NLTE levels} & lines & \\
    \noalign{\smallskip}
    \hline
    \noalign{\smallskip}
    H  & \textsc{i}   &  10 & & 36 & \\
    \noalign{\smallskip}
    He & \textsc{i}   &   5 & &  3 & \\
       & \textsc{ii}  &  10 & & 36 & \\
    \noalign{\smallskip}
    C  & \textsc{iii} &   6 & &  4 & \\
       & \textsc{iv}  &   4 & &  1 & \\
    \noalign{\smallskip}
    N  & \textsc{iv}  &   6 & &  4 & \\
       & \textsc{v}   &   4 & &  1 & \\
    \noalign{\smallskip}
    O  & \textsc{iv}  &   1 & &  0 & \\
       & \textsc{v}   &   6 & &  4 & \\
       & \textsc{vi}  &   4 & &  1 & \\
    \noalign{\smallskip}
    Fe & \textsc{iv}   & 7 & (6\,472) & 25 & (1\,027\,793)  \\
       & \textsc{v}    & 7 & (6\,179) & 25 &    (793\,718)  \\
       & \textsc{vi}   & 8 & (3\,137) & 33 &    (340\,132)  \\
       & \textsc{vii}  & 9 & (1\,195) & 39 &     (86\,504)  \\
       & \textsc{viii} & 7 &    (310) & 27 &      (8\,724)  \\
    \noalign{\smallskip}
    Ni & \textsc{iv}   & 7 & (5\,514) & 25 &    (949\,506) \\
       & \textsc{v}    & 7 & (5\,960) & 22 & (1\,006\,189) \\
       & \textsc{vi}   & 7 & (9\,988) & 22 & (1\,110\,584) \\
       & \textsc{vii}  & 7 & (6\,686) & 18 &    (688\,355) \\
       & \textsc{viii} & 6 & (3\,600) & 20 &    (553\,549) \\
    \noalign{\smallskip}
    \hline
    \noalign{\smallskip}
    total   &     & 135 & (49\,041) & 346 & (6\,565\,054)\\
    \noalign{\smallskip}
    \hline
  \end{tabular}

\end{table}

\begin{figure*}[htp]
  \resizebox{\hsize}{!}{\includegraphics{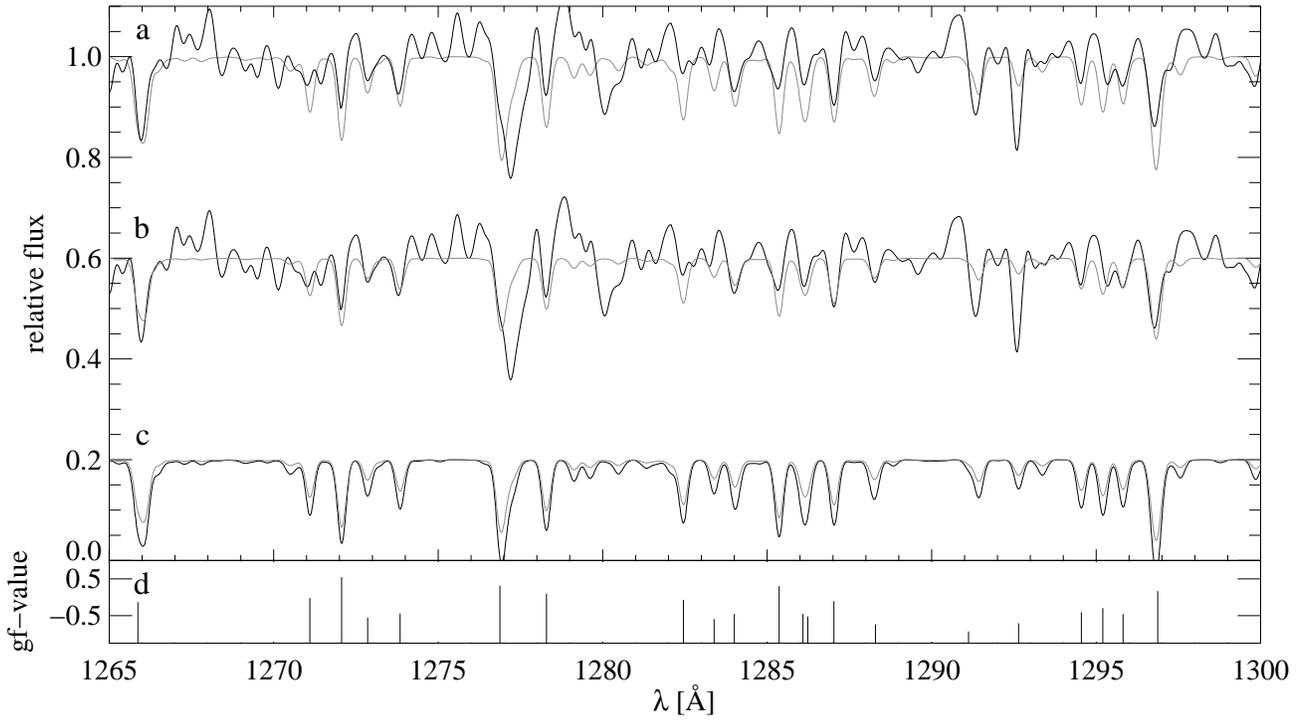}}   
  \resizebox{\hsize}{!}{\includegraphics{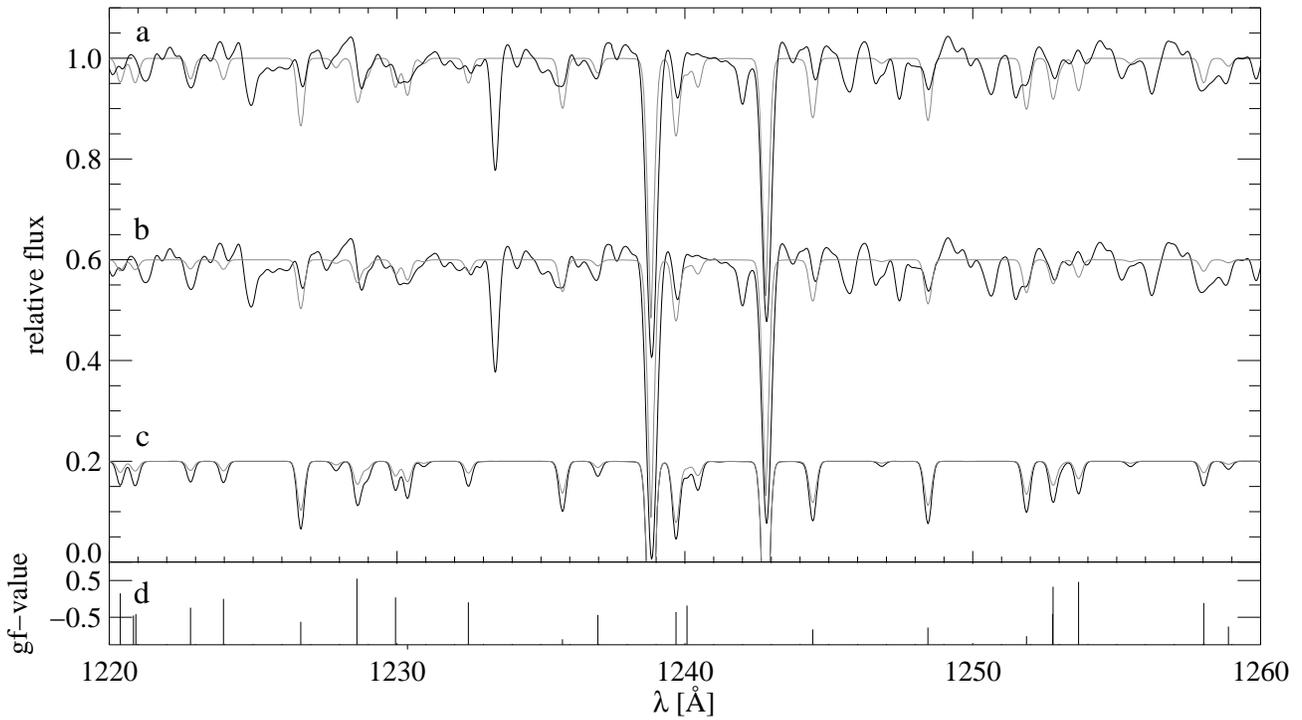}}   
  \caption{Details from fits to the iron line spectrum of two central stars.
    \textbf{Top:} S\,216; $T_{\rm eff}$ = 90\,000 K, $\log g$ = 7.0;
    \textbf{Bottom:} NGC\,1360; $T_{\rm eff}$ = 110\,000 K, $\log
    g$ = 6.0.
    Models (dotted):
    \textbf{a} solar Fe and Ni;
    \textbf{b} 0.5~dex subsolar Fe and Ni; 
    \textbf{c} comparison of both models; 
    \textbf{d} strong \ion{Fe}{vi}
    and \ion{Fe}{vii} lines are marked.
  }
  \label{fig:spectra}
\end{figure*}

\begin{figure*}[htp]  
  \resizebox{\hsize}{!}{\includegraphics[width=0.5\textwidth]{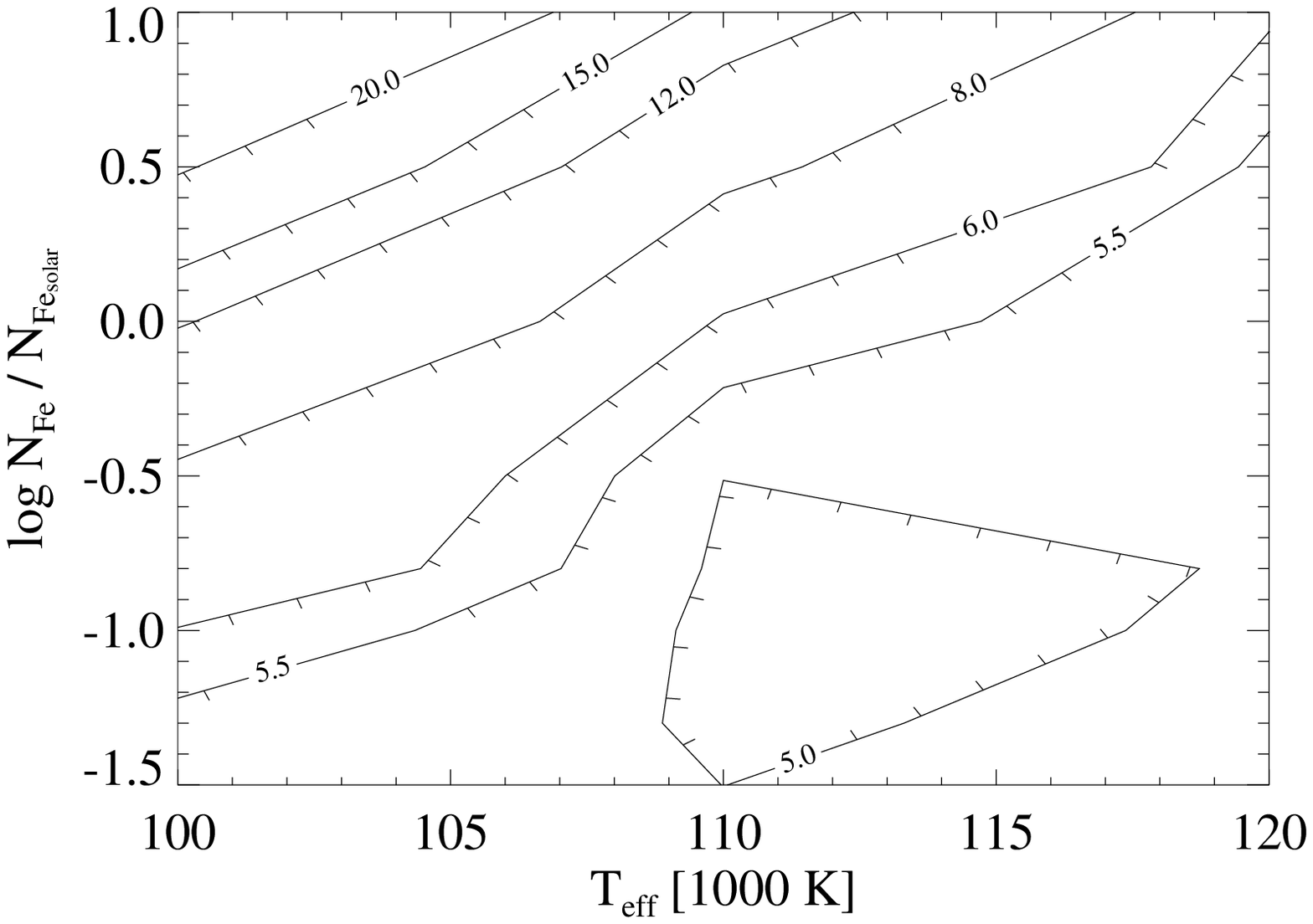}
    \includegraphics[width=0.5\textwidth]{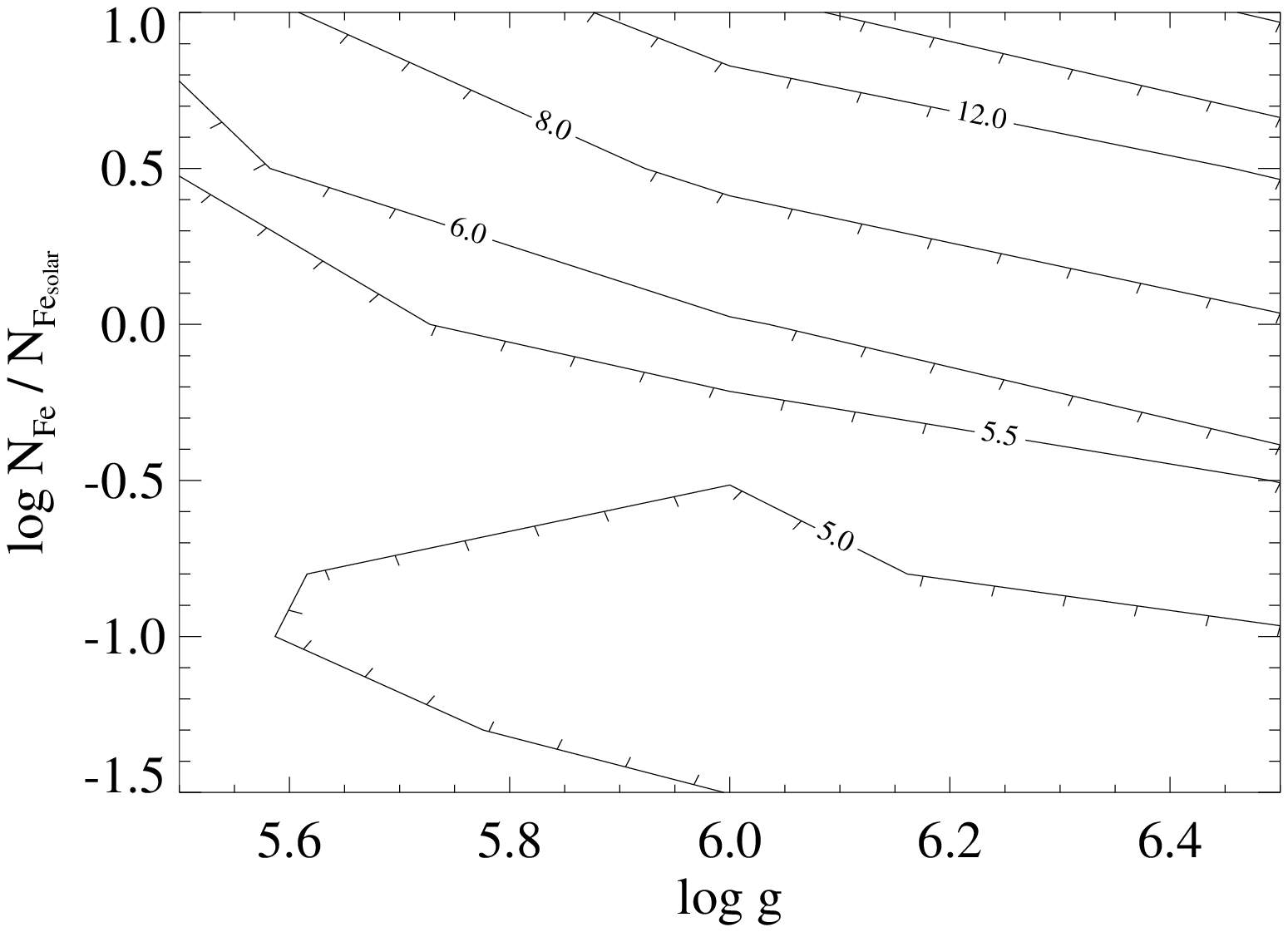}}
  \caption{Contour plot of the $\chi^{2}$ test for the
    spectrum of NGC\,1360. In the left plot the $\chi^{2}_{\rm red}$ values
    for those models with $\log g$ at 6.0 and $T_{\rm eff}$ around
    110\,000\,K ($40 = 8 \cdot 5$ models, see Fig.~\ref{fig:grid}) are
    shown, whereas in the right plot $T_{\rm eff}$ remains fixed and $\log
    g$ varies. The high $\chi^{2}$ values ($> 5$) are due to the low S/N
    level and the imperfect normalization (see Fig.~\ref{fig:continuum}).
    As a general trend, the $\chi^{2}$ test favors models with subsolar
    iron abundance}
  \label{fig:chisq}
\end{figure*}

\section{Results and discussion}
We derived our results with two independent methods. First we compared the
observed and theoretical spectra by eye (Fig.~\ref{fig:spectra}), second a
$\chi^{2}$-test has been performed (Fig.~\ref{fig:chisq}). To get more
reliable results, the $\chi^{2}_{\rm red}$ has been computed only around
strong iron lines excluding known ISM lines. As a general trend, the iron
abundance appears to be subsolar by 0.5-1~dex with both methods (see
Tab.~\ref{tab:objects}). However, the S/N of the IUE spectra is not
sufficient to exclude a solar abundance in any specific case.

We cannot identify nickel lines for certain, which correspond to an upper
limit of a solar nickel abundance. \ion{Ni}{v} lines have been detected in
other central stars and sdO stars by \cite{schoenberner:85a}, however,
these stars have effective temperatures which are lower than those of our
objects.

For three high-gravity objects in the sample the suspected iron deficiency
may be the result of gravitational settling. For the other stars of the
sample with low gravity and high luminosity another explanation must be
found since mass loss efficiently suppresses gravitational settling
\citep{unglaub:98a}. Following suggestions in the literature
\citep{lambert:88a, venn:90a, bond:91a, winckel:92a}, the
iron abundance in these stars might not be primordial, but the result of
chemical fractionation through dust formation and subsequent loss due to
stellar wind during previous AGB phases.

Therefore improved spectroscopy either by FUSE or HST is necessary to
confirm or reject the possibility of a general Fe deficiency in these
central stars. Furthermore, the determination of the abundance of
CNO-elements is desirable in order to get a better understanding of the
chemical evolution of these stars.

Our study suggests that iron is not necessarily a suitable tracer to
determine the primordial metallicity. Alternatively one may use the S or Zn
abundance to obtain reliable information \citep{winckel:92a}.  The
abundance of both elements is not affected by nuclear processes during
previous evolutionary phases and both elements do not tend to dust
formation as iron does \citep{habing:96a}. Due to the lack of detailed
atomic data for Zn, one has to concentrate on the S lines in the FUV and UV
region.  This emphasizes the need for improved spectroscopic data to solve
these questions.

%
\begin{acknowledgements}
  We thank S.~Haas for helpful discussions and for providing his original
version of the iron cross-section sampling software. UV data analysis in
T\"ubingen is supported by the DLR under grants 50\,QV\,97054 and 50\,OR\,97055.

\end{acknowledgements}

%

%
\end{document}